\documentclass[12pt,onecolumn]{bmcart}

\usepackage{amsthm,amsmath}
\usepackage{amsmath,amssymb}
\usepackage{graphicx}

\usepackage{color}

\newcommand{\ket}[1]{\vert #1 \rangle}
\newcommand{\bra}[1]{\langle #1 \vert}

\begin{document}

\begin{frontmatter}

\begin{fmbox}


\title{Quantum quenches and thermalization on scale-free graphs} 


\author{Francesco Caravelli}


\address[id=aff1]{
  \orgname{QASER Lab, University College London}, 
  \street{Gower Street},                     %
  \city{London},                              
  \cny{UK}                                    
}


\end{fmbox}


\begin{abstractbox}
\begin{abstract} 
We show that after a quantum quench of the parameter controlling the number of particles in a Fermi-Hubbard model on scale free graphs, the distribution of energy modes follows a power law dependent on the quenched parameter and the connectivity of the graph. This paper contributes to the literature of quantum quenches on lattices, in which, for many integrable lattice models the distribution of modes after a quench thermalizes to a Generalized Gibbs Ensemble; this paper provides another example of distribution which can arise after relaxation.  We argue that the main role is played by the symmetry of the underlying lattice which, in the case we study, is scale free, and to the distortion in the density of modes.
\end{abstract}


\begin{keyword}
\kwd{Quantum quenches,}
\kwd{scale free,}
\kwd{spectrum density,}
\kwd{Generalized Gibbs Ensemble}
\end{keyword}


\end{abstractbox}
%

\end{frontmatter}


\begin{backmatter}

\normalsize

\section{Introduction}
There has been recent interest in the effective thermal dynamics following a quantum quench in spin chains \cite{Rossinietal}. The dynamics out of equilibrium of quantum systems \cite{fransson} has received a great amount of attention \cite{Belgiorno,Cacciatori:2010vr,Rubino:2011zq,Calabreseetal}. It became clear that after a quantum quench, many observables at equilibrium after a quantum quench are distributed according to a Generalized Gibbs Ensemble (GGE) \cite{Rigol,Rossinietal,Calabreseetal,Fagotti} or a Gibbs ensemble. In addition to these theoretical understandings of the thermalization in quantum systems, these results are supported by recent experiments in trapped cold atomic gases\cite{experiments}.
Many of these experiments focused on the role played by dimensionality and conservation laws, which in turn initiated a vigorous effort in understanding the role of the integrability of the system under scrutiny at late times. In fact, the easiest way of driving a quantum system out of equilibrium is indeed a \textit{quantum quench} (see for instance \cite{polr} for a comprehensive summary),
i.e. a sudden change in the parameters of a Hamiltonian, and its subsequent relaxation at long times. It has been argued that for integrable models many observable are distributed according to a Generalized Gibbs Ensemble, meanwhile an effective Gibbs distribution arises in generic systems. Recent studies suggest that the behavior is indeed more complicated, showing a dependence on the initial conditions (state)\cite{state}.

In a previous work, we studied quantum quenches in a Fermi-Hubbard model which does not conserve the number of particles \cite{Quenchesus}. We studied the energy of the excitations, which are invariant under time evolution after the quench, and found that these are distributed according to a GGE.  There, the temperature is associated with the gap in the spectrum, which is due to the coupling of non-conserving number of particles term. A similar phenomenon happens in quantum liquids \cite{Leggett}. 

In this paper we explore a similar approach on a different type of underlying interaction network. Several classical statistical models have been studied on complex networks \cite{CPcn}.
Complex networks have become an area of tremendous recent interest since the discoveries of the small world and scale free properties in many realistic networks. A small world network is characterized by short network distance a high clustering coefficient. Several reviews of the subjects are now available \cite{barev,caldarelli,dyncn}. 
Important applications of these techniques are spreading of diseases \cite{escn} and syncronization\cite{synccn} on complex networks. Watts and Strogatz demonstrated that the two small-world characteristics can be obtained from a regular network by
rewiring or adding a few long-range links shortcuts, which connect otherwise distant nodes \cite{ws}. A regular network intrinsically already has a high clustering coefficient, and has a diameter which is logarithmic in the number of nodes. However, few shortcuts can reduce the distance exponentially unaffecting the clustering coefficient, but dramatically reducing the average diameter of the graph.  
The scale-free property is characterized by an algebraic degree distribution:  where the degree variable k measures the number of links of node in the network, $P(d\gg1)\approx d^\gamma$, and where $\gamma$ is the algebraic scaling exponent. Barab\'{a}si and Albert discovered the scale-free property and also proposed growth and preferential attachment as the two basic mechanisms responsible for the scale-free property. Here, growth requires that the numbers of nodes and links increase with time and preferential attachment means that when a new node is added to the network, the probability that it connects to an existing node is proportional to the number of links that this node has already had.
In this paper, we consider the properties of quantum quenches on Barab\'{a}si-Albert type of graphs. It is important to understand what kind of distributions can arise in a quantum quench. Although \textit{per se} the study of quantum phenomena on complex networks might not be physically relevant, it is important from the theoretical point of view. While GGE is a quite common example of distribution arising, here we show that another type of distribution arise.
This paper is organized as follows. In section 2 we introduce Quantum Quenches, Scale Free graphs and the Model. In section 3 we describe the results, meanwhile Conclusions follow in section 4.

\begin{figure}
\centering
\includegraphics[scale=0.5]{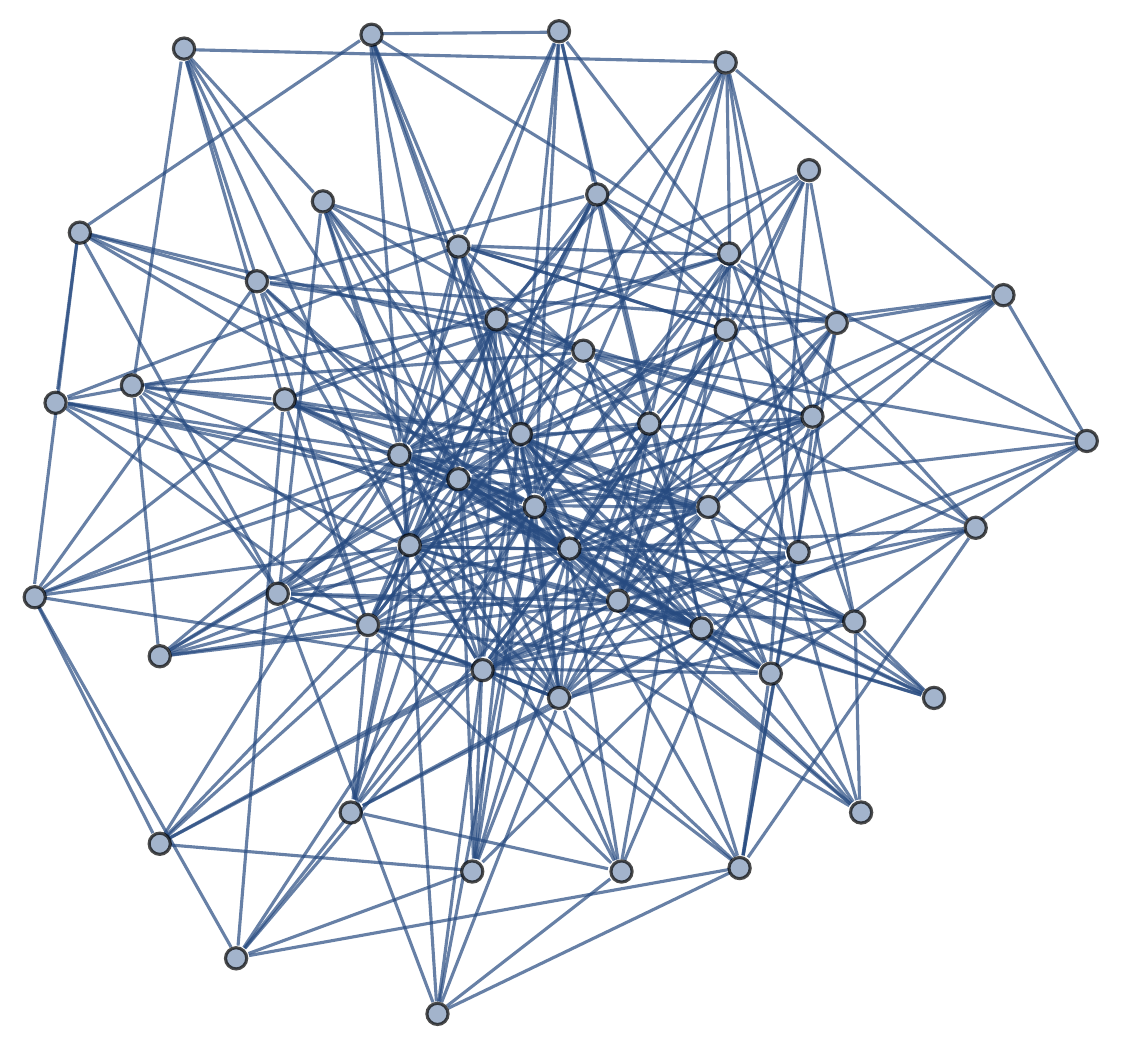}
\caption{An example of Barab\'{a}si-Albert graph.}
\end{figure}

\section{The model}

\subsection{Quantum Quenches and GGE}

In this section we provide a more detailed introduction to thermalization after a quantum quench, and introduce the Generalized Gibbs Ensemble which we will later discuss of.  Ergodicity in a classical statistical physics setting accounts for the independence of the asymptotic state distribution from the initial condition: that is, at large times thermal equilibrium is approached. Meanwhile in classical mechanics thermalization occurs thanks to ergodicity, in quantum
mechanics of isolated systems the unitary dynamics is an obstruction to obtaining an anagolous result: if the system is initially prepared in a pure state it will remain in a pure state, as the evolution is period or quasiperiodic, which means that after a sufficient long time it will return to the initial state. 
However, one can focus on certain expectation values only or trace out some part of the system, i.e. subsystems of the whole are not isolated, and thus the reduced density matrix is not pure anymore. In general, the type of question one asks in thermalization of quantum system is: how close are expectation value in a subsystem, to those of the same degrees of freedom averaged with a Gibbs ensemble? In general, one does not get exactly a Gibbs ensemble, but a Generalized Gibbs Ensemble:
\begin{equation}
\rho_{GGE}\approx e^{-\sum_j \lambda_j I_j}
\label{eq:GGEc}
\end{equation}
in which one introduced a Lagrange multiplier $\lambda_j$ and $I_j$ are the integral of motion, which generalizes the standard Gibbs distribution. The GGE-conjecture states that the stationary expectation value of any local observable are equal to the ensemble expectation values or  equivalently that of the reduced density matrix of local observables. The subtetly is that one does not have a recipe for choosing the integrals of motion in eqn. (\ref{eq:GGEc}). The general approach to study thermalization is through the device of quantum quenches: you change suddendly a parameter and observe how the system thermalizes. The GGE-hypothesis has been tested extensively for many systems, and has turned out be valid for many quantum quench problems stuedied recently. In general, this is true for non-interacting models, or models that can be casted into a non-interactive one (quadratic). In general, the mathematical device used to prove that the GGE is valid is the Wick theorem, and that the initial state overlap with the post-quench quasi-particle modes is gaussian. An underlying assumption is thus that different modes are orthogonal, which is crucial in order to prove Wick's theorem. In fact, if the underlying lattice is translational invariant, one obtains that two different modes, in general of the form $e^{i k x}$, are orthogonal, e.g. $\langle \phi_{k_1}(x),\phi_{k_2}(x)\rangle=\int  e^{i(k_1-k_2) x} dx=\delta(k_1-k_2)$. In this work we  consider the case in which the expansion does not have such a clear interpretation in terms of momentum, which occurs in the case in which translational invariance is not explicit in the underlying lattice. However, we consider the case in which the graph has an overall symmetry, i.e. the distribution of the degree is well approximated by a power law for large values, and thus is scale invariant in its tail, which will be introduced in the following section. In addition to this, the density of energy modes is, differently from the case of translational invariant graphs, distributed according to a power law.

\subsection{Scale-free graphs}

We now recall the growth algorithm used in the preferential attachment model introduced by Barab\'{a}si and Albert \cite{ba}.
The growth algorithm (preferential attachment) is parametrized by a single parameter, $M$. The starting graph is a single node, with no edges. Then, 
at each step, a new node is added, with $M$ edges. The edges are attached at random to the previous existing nodes, with a probability proportional to the degree of the node.
If $d_i$ is the degree of the vertex $i$, at each edge is attached to a node i with probability $p_i=\frac{d_i}{\sum_i d_i}$. As it is well known, these graphs are scale-free, i.e.  for $N\gg1$, 
the degree distribution is a power-law, $P(d\gg1)\approx d^\alpha$, with $\alpha$ exponent of the power law. Another property, here important to mention, is that scale-free graphs are \textit{ultra-small}: the average distance between two nodes goes as $\sim\log(\log(N))$, where $N$ is the number of nodes in the network. Notably, Bose-Einstein condensation appear in growing networks if preferential attachment growth is generalized with fitness\cite{bianbar}.

\subsection{Hamiltonian and Quantum Quench}
Here we want to recollect the formalism introduced in \cite{Quenchesus}. The Hamiltonian we will consider in the present paper is the following Fermi--Hubbard model:

\begin{eqnarray}
H(\Gamma_M,\lambda)=&-&J\sum_{i,j=1}^{N_v} A_{ij}^{(\Gamma_M)}a^\dagger_ia_j \nonumber \\
&+&\frac{\lambda }{2}\sum_{i,j=1}^{N_v} B_{ij}^{(\Gamma_M)} \left(a^\dagger_ia^\dagger_j+\textrm{h.c.}\right),
\label{eq:full-hamiltonian}
\end{eqnarray}
where $a_i$ ($a^\dagger_i$) is the annihilation (creation) fermionic operator that annihilate (create) a particle in the vertex $i$
of the background graph $\Gamma_M$. 
The matrices $A_{ij}^{(\Gamma_M)}$ and $B_{ij}^{(\Gamma_M)}$ are, respectively, the adjacency matrix of $\Gamma_M$
and its antisymmetrized form. In the present paper, the adjacency matrix will be the one of a scale-free graph built using the Barab\'{a}si-Albert growth
algorithm. The sum runs over all the $N$ nodes of the graph $\Gamma_M$, where $M$ is the connectivity parameter introduced previously.
The coupling $J$ is the tunneling of the particles between two connected sites and $\lambda$ controls the strength of the Hamiltonian terms
that do not conserve the number of particles. The physical properties are independent from time-scaling if we perform a sudden quantum quench, thus we can measure
excitations in units of $J$\footnote{ Since now on we set $J=1$ and measure $\lambda$ in units of $J$.}.

In particular, we introduced a notion of particle (with an associated discrete labeling $k$, 
$\epsilon(k)$) given in terms of ladder operators $\eta_k$.  
Once the notion of particle that the detector measures is established, we
can determine the energy distribution (number of particles with momentum $k$) of the ground state of the system 
\begin{equation}
n(k) = \bra{GS} \eta_k^\dagger \eta_k \ket{GS}.
\label{eq:momentum-distribution}
\end{equation}
For graphs with discrete translational invariance, this is associated with the Fourier transform vectors $e^{i k x}$, but for graphs without particular symmetry this
identification is lost.

The notion of particle $\eta_k$, together with its dispersion relation
$\epsilon(k)$ will be defined in terms of a test Hamiltonian 
\begin{equation}
H_{\mathrm{test}}=\sum_k \epsilon(k) \eta_k^\dagger \eta_k \, .
\end{equation}
The momentum distribution \eqref{eq:momentum-distribution} that the detector measures is 
given by the overlap between the ground state of the system and the eigenstates states of the test Hamiltonian $H_{\mathrm{test}}$.

The quantum quench we are going to perform is given by
$$H(\Gamma_M,\lambda)\rightarrow H(\Gamma_M,0),$$
from the ground state of $H(\Gamma_M,\lambda)$, $|GS\rangle$,.
 After the quench. The system $H(\Gamma_M,0)$ will see the ground state $|GS\rangle$ as an excited state, and thus it makes sense to calculate the spectrum-density $n(\epsilon(k))$. 
The hopping Hamiltonian can be written as
\begin{equation}
H_{\mathrm{test}} = \sum_{i,j=1}^{N_v} A_{ij}^{(\Gamma_M)} a_i^\dagger a_j=\sum_{k=1}^{N_v} \epsilon(k)\eta^\dagger_{k}\eta_{k}.
\label{eq:H-detector}
\end{equation}
The eigenmodes of $H_{\mathrm{test}}$, labeled by an integer $k$, and with energy $\epsilon(k)$, define our notion
of particle. These are created and annihilated by the operators $\eta_k^\dagger$ and $\eta_k$, 
and are the excitations that the detector measures, and that we will calculate.
Therefore, we need to compute 
\begin{equation}
n(k) = \bra{GS} \eta^\dagger_{k}\eta_{k} \ket{GS}\, ,
\label{counts}
\end{equation} 
and calculate the distribution. As we will see, $n(k)\approx \epsilon(k)^\gamma$, where $\gamma$ is the exponent we will study. 
The two Hamiltonians have the same number of nodes, thus their Hilbert states overlap (coincident). 
\footnote{\footnotesize The Hamiltonian \eqref{eq:full-hamiltonian} is a quadratic model, hence, it can be diagonalized as
\begin{equation}
H = \sum_{q=0}^{N_v} \omega(q) \psi_{q}^\dagger \psi_{q},
\end{equation}
by means of a Bogoliubov transformation of the fundamental particle operators, $a_{i}^\dagger,a_{j}$.
In turn, these are related by another Bogoliubov transformations to the operators
$\eta,\eta^\dagger$. Then, the operators $\eta, \eta^\dagger$ will be connected to the $\psi,\psi^\dagger$ by
the Bogoliubov transformation that is the composition of the Bogoliubov transformations that
relate  $\psi,\psi^\dagger$ to $a,a^\dagger$ and $a, a^\dagger$ to $\eta, \eta^\dagger$.
It can be written formally as
 \begin{equation}
\eta_{k} = \sum_{q=0}^{N_v} \left( \alpha_{kq} \psi_{q} + \beta_{kq} \psi^\dagger_{q} \right)\, ,
\end{equation}
where $\alpha_{kq}$ and $\beta_{kq}$ are the Bogoliubov coefficents.}
If the system were at equilibrium with an external bath at temperature $T$, we would have 
$n(k)=e^{-\frac{\epsilon(k)}{T}}$. 
In the case of an $n$-dimensional tori, we found in \cite{Quenchesus} for the same model we study here, that
$n(k)=e^{-\frac{2 \epsilon(k)}{\lambda}}$,  where $\lambda/2$ plays the role of the temperature. 

The spectrum of the adjacency matrix of a scale-free network was studied in \cite{spectraba}.
It is known that eigenvalues are distributed according to a power law, while for the eigenvectors an analytical form is still lacking.
However, the components of the eigenvectors are strongly localized at the hubs. 

As a matter of fact, in our numerical calculation we will neglect the role of the shape of the Fermi surface, approximating it with a sphere. 
This implies that a small error is made on the temperature, that can become more and more relevant for $\lambda\approx 1$. Thus, our result is valid in the limit $|\lambda|\ll 1$.

\section{Results}

Here we show that the distribution of the modes in the ground state follows a power law.
Since analytical techniques are lacking, we evaluated eqn. (\ref{counts}) numerically. all the numerical results are obtained 
with a number of nodes $N=700$.  We can tackle such a big quantum system, due to the Bogoliubov transformation which diagonalizes this model; the quantum dynamics of this system is in fact restricted to a Hilbert space of size $2N$, instead of the $2^N$ of an ordinary spin system. In Fig. \ref{loglog} we plot  $\log(n(k))$ against $\log(\epsilon(k))$. The functional dependence appears, numerically, to be of the form:
\begin{equation}
n(k)\sim f(M,\lambda) \epsilon(k)^{\gamma}
\end{equation}
where $\gamma$ is weakly dependent on $\lambda$, in the range $\lambda\in[0,0.2]$ and takes value $\gamma\in [ -0.85,-0.95]$.
Tables \ref{tab:table1} and \ref{tab:table2} show the values obtained numerically by fixing $M$ and $\lambda$ and fitting the power law, together with the error,
of the parameters $f$ and $\gamma$.  In Fig. \ref{fgamma} and Fig. \ref{fM} we plot the functional dependence of the constant in front of the Zipf's law, 
by fixing the values of $M$ and $\lambda$. These appear to be both convex functions of the parameters.
The power law is rather stable over several order of magnitudes, $\log(\epsilon(k))\in[-2,2]$, although the density of points
is higher in $\log(\epsilon(k))\gtrsim 0$.
For $\lambda\in[0.2,1]$, the distribution changes shape in $\log(\epsilon(k))<0)$. Thus, restoring the units, the power law for density of excitations is valid only in the limit $\frac{\lambda}{J}\ll 1$.
It is clear from this analysis that the power law exponent, in its domain of validity, is independent from the connectivity parameter $M$.

One can give a rough explaination of the results above, by considering the exact expression of $n(\epsilon)=\frac{\lambda^2(1-\epsilon^2)}{\lambda^2(1-\epsilon^2)+\epsilon^2+\epsilon\sqrt{\lambda^2(1-\epsilon^2)+\epsilon^2}}$ \cite{Quenchesus}.
For $\lambda\ll 1$, one has that $n(\epsilon)\approx \frac{\lambda}{2}(\frac{1}{\epsilon^2}-1)$.
However, if the density of eigenvalue $\rho(\epsilon)$ is distributed according to a power law, one has to smear the effective distribution considering the deformed density. In fact, if we consider
$\langle n(\epsilon) \rangle=\int n(\epsilon) \rho(\epsilon) d\epsilon\approx \frac{1}{\epsilon^{1+\alpha}}$, depending on the right functional, asymptotic expression of $\rho(\epsilon)$, and considering that usually the function $\rho(\epsilon)$ is peaked at $\log(\epsilon)\approx 0$, which is what we observe numerically.
This should explain why we observe a power law for small values of $\lambda$.

\begin{figure}
\centering
\includegraphics[scale=0.7]{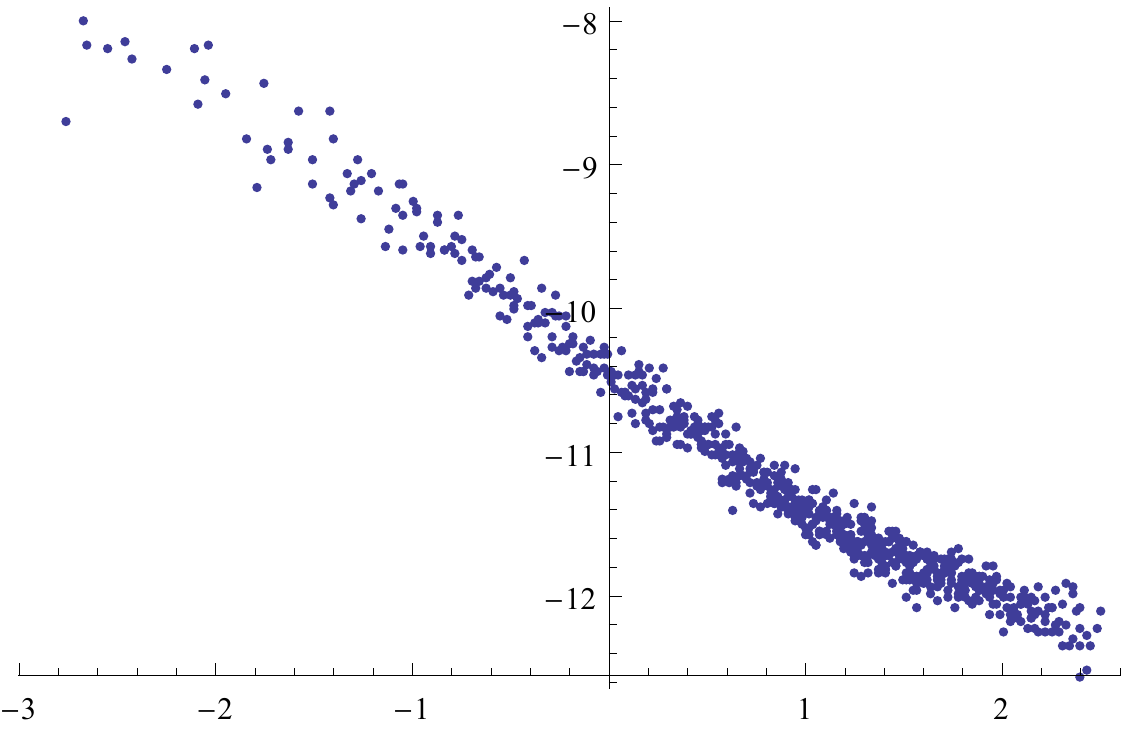}
\caption{Log-Log plot of the distribution $n(\epsilon(k))$, evaluated for $N=700$, $M=10$, $\gamma=0.1$. $\log(n)$ on the y-axis, $\log(\epsilon)$ on the x-axis.}
\label{loglog}
\end{figure}

\begin{figure}[b]
\centering
\includegraphics[scale=0.9]{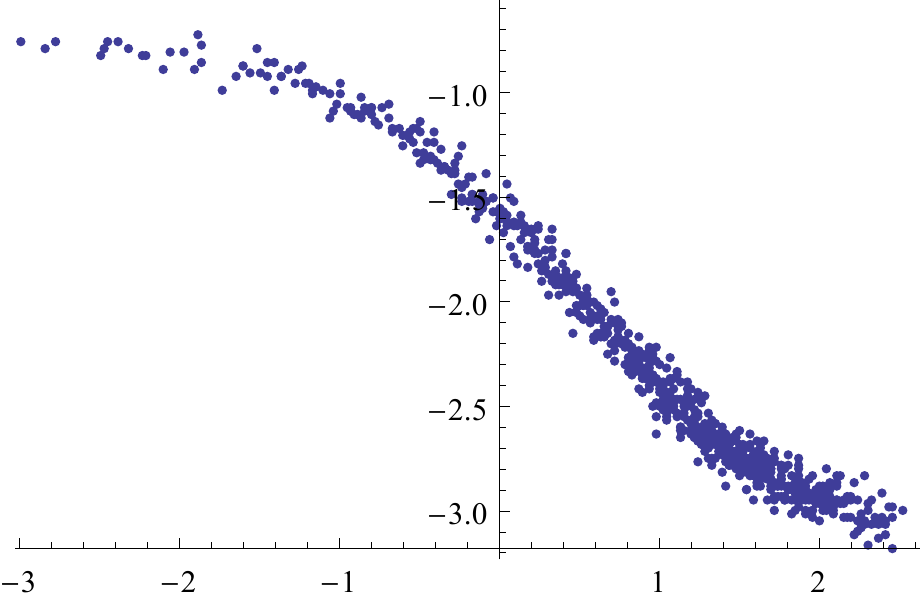}
\caption{Log-Log plot of the distribution $n(k)$ for  $M=10$., $\lambda=0.8$ and $N=700$.}
\label{populationrest}
\end{figure}

\section{Conclusions}

In this paper we discussed quantum quenches of a Fermi-Hubbard model on scale-free graphs, motivated by the search to alternative distributions from those of the Generalized Gibbs Ensemble hypothesis, relevant in particular in the case of integrable lattice models.
A previous analysis of the quench protocol discussed in the present paper was done in \cite{Quenchesus}, and solved analytically for the case of $n$-dimensional torii, showing that the Generalized Gibbs Ensemble emerges (GGE). The quenched parameter, $\lambda$, controls the conservation of the particle number and, in particular, introduces a gap in the spectrum. For non-integrable systems, it is a known fact that the expectation value of several observables after a quantum quench, are similar to those calculated on a thermal state, i.e. a GGE. The observables we considered were the density of eigenmodes calculated over the ground state of the unquenched Hamiltonian. In this work, we have shown that the spectrum of the excitations for scale free graphs can be well approximated by a power law for the case in which the quenched parameter is small. In particular, we have analyzed the functional dependence of the two parameters of the power law on the topological properties of the scale-free graph, the connectivity, and the quenched parameter. We found that while small values of $\lambda$ this distribution is scale free, meanwhile for higher values of the quenched parameter ($\lambda>0.2$), the distribution changes shape and the power law behavior is lost.

The outcome of our analysis shows that the underlying symmetry of the graph does indeed contribute the to the shape of the distribution of modes. While GGE naturally emerges in the case of translational invariant lattices, we have given evidences for a counter example, in which a power law distribution for the modes spectral density arises. 
Due to the lack of analytical understanding of the eigenmodes for scale free graphs, we have reached these conclusions using a numerical approach. We have focused on a Fermi-Hubbard model for a quite simple, technical reason. The scale free properties of graphs generated using a preferential attachment become evident for large graphs, e.g. for the case $N\rightarrow \infty$. We thus had to focus on a model which can be diagonalized numerically for large graphs. For the Fermi-Hubbard model considered in the present paper, we could take advantage of a Bogoliubov transformation which allows to tackle the diagonalization process on a matrix of size proportional to the number of nodes, and not exponentially increasing, as standard in quantum mechanics. 

We have tried to show in a numerically treatable model the relevance of the symmetry of the underlying lattice for the GGE hypothesis. Although striking evidences have been put forward, both analytically and numerically, of the universality of the GGE for 
integrable models \cite{Calabreseetal}, these results rely on the underlying symmetry of the graph, in particular on the properties of the eigenvectors of the adjacency matrix of the lattice. These properties are indeed lost when translational invariance is not present, which is the case for the scale free graphs. However, we have shown that the distribution does indeed retain some properties of the underlying lattice, which for the case of scale free graphs is the scale-freeness of the spectrum distribution. We have provided a rough, analytical explaination of why, in the case in which the density of modes is not uniform, one has to consider a smeared version of the spectral density.

\section*{Acknowledgements}
{\small We would like to thank Arnau Riera and Lorenzo Sindoni for extensive collaboration on this subject.}

\clearpage
\newpage
\appendix
\section{Supplementary Material}
\begin{table}[h]
\centering
\begin{tabular}{|c|c|c|c|c|}
\hline
M & $f$ & $\Delta f$ & $\gamma$ & $\Delta \gamma$ \\
\hline
4 & -7.760 & 0.006 & -0.809 &  0.006 \\
5 & -7.591 & 0.007 & -0.843 &  0.007\\ 
6 & -7.528 & 0.007 & -0.837 & 0.006\\
7 & -7.404 &  0.007 & -0.852 & 0.005 \\
8 & -7.332 &  0.008 & -0.857 & 0.006\\
9 & -7.255 & 0.008 & -0.859 & 0.006\\
10 & -7.140 &  0.008 &-0.905 & 0.006\\
11 & -7.036 &  0.009 & -0.942& 0.006\\
12 & -7.071 &  0.008 & 0.884  & 0.006\\
13 & -6.982 &  0.008 &  -0.924 & 0.006\\
14 & -6.978 & 0.008 & -0.888 & 0.006\\
15 & -6.882 &  0.009 & -0.927 & 0.006\\
16 & -6.871 & 0.009 & -0.917 & 0.006\\
17 & -6.909 & 0.009 & -0.891 & 0.006\\
18 & -6.820 &  0.009& -0.919 &  0.005\\
19 & -6.801 & 0.009 & -0.911 & 0.005\\
20 & -6.83 & 0.01& -0.877 & 0.006 \\
\hline
\end{tabular}
\caption{Table of $\gamma$ and $f$ as function of $M$, for fixed $\lambda=0.05$ and $N=700$.}
\label{tab:table1}
\end{table}

\begin{table}[h]
\centering
\begin{tabular}{|c|c|c|c|c|}
\hline
$\lambda$ & $f$ & $\Delta f$ & $\gamma$ & $\Delta \gamma$ \\
\hline
0.01 & -10.407 & 0.008 & -0.891 & 0.006 \\
0.02 & -9.011 &   0.008 &  -0.887 & 0.006 \\
0.03 & -8.166 & 0.008 &  -0.896 & 0.006 \\
0.04 & -7.656 &  0.008 & -0.869 & 0.005 \\
0.05 & -7.171 & 0.008 & -0.880 & 0.005 \\ 
0.06 & -6.810 & 0.008 & -0.884 &  0.005 \\
0.07 & -6.483 & 0.007 &  -0.889 & 0.005 \\
0.08 & -6.213 & 0.008 & -0.901 & 0.005 \\
0.09 & -5.988 & 0.008 &  -0.889 & 0.005 \\
0.1  &  -5.810 & 0.007 & -0.876 & 0.005 \\
0.11 & -5.577 & 0.007 & -0.897 & 0.005 \\
0.12 & -5.455 & 0.008 & -0.870 & 0.005 \\
0.13 & -5.283 & 0.008 & 0.866 & 0.005 \\
0.14 & -5.121 & 0.008 & -0.880 & 0.005 \\
0.15 & -4.985 & 0.007 & -0.882 & 0.005 \\
0.16 & -4.861 &  0.008 & -0.876 & 0.005 \\
0.17 &-4.7286 &   0.007 & -0.873 & 0.005 \\
0.18 & -4.569 &   0.008 & -0.902 & 0.006 \\
0.19 & -4.519 &   0.007 & -0.874 & 0.005 \\
0.2  & -4.390 &  0.008 & -0.870  & 0.005 \\
\hline
\end{tabular}
\caption{Table of $\gamma$ and $f$ as function of $\lambda$, for fixed $M=10$ and $N=700$.}
\label{tab:table2}
\end{table}
\clearpage
\newpage
\begin{figure}
\centering
\includegraphics[scale=1.05]{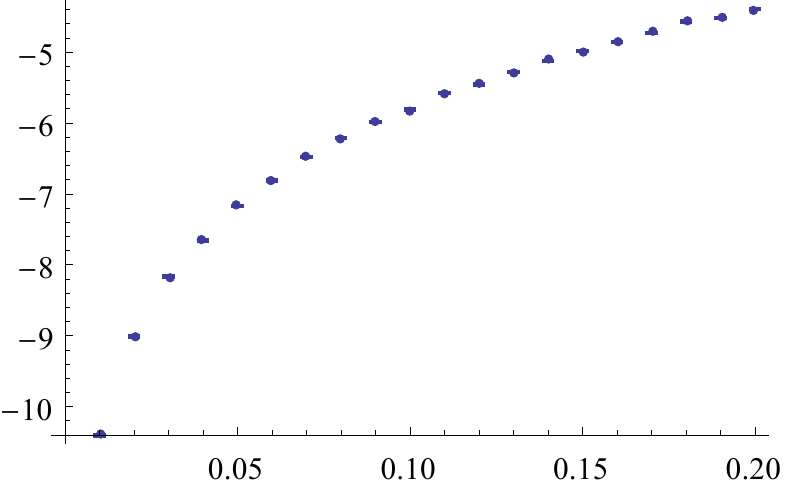}
\caption{Plot of the $f$ as a function of $\gamma$ for $M=10$ and $N=700$.}
\label{fgamma}
\end{figure}

\begin{figure}
\centering
\includegraphics[scale=0.85]{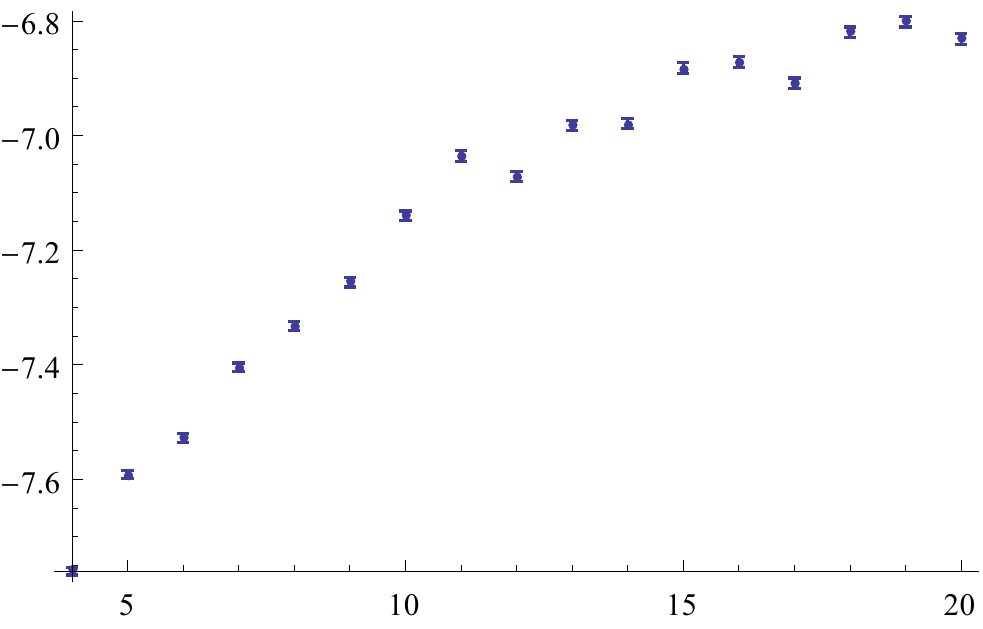}
\caption{Plot of $f$ as a function of $M$, for $\lambda=0.05$ and $N=700$.}
\label{fM}
\end{figure}

\end{backmatter}
\end{document}